\documentclass[aps,prl,twocolumn,superscriptaddress,amsmath,amssymb,longbibliography]{revtex4-2}
\usepackage[colorlinks=true, linkcolor=blue, anchorcolor=blue, citecolor=blue, urlcolor=blue]{hyperref}

\usepackage{etoolbox}
\usepackage{graphicx}  
\usepackage{epstopdf}
\usepackage{dcolumn}   
\usepackage{bm}        
\usepackage{amssymb}   
\usepackage{amsmath}   
\usepackage{amsthm}
\usepackage{chemarrow}
\usepackage{color}
\usepackage{mathrsfs}
\usepackage{float}
\usepackage{bbold}
\usepackage{bbm}
\usepackage{physics}
\usepackage{cancel}
\usepackage{siunitx}
\usepackage{cleveref}

\begin{document}
	
	\title{Temporal Dynamical Quantum Phase Transition in Dicke Model with Trapped Ions}

    \author{Ji Bian}
\thanks{These authors contributed equally to this work.}
\affiliation{Laboratory of Spin Magnetic Resonance, School of Physical Sciences, Anhui Provincial Key Laboratory of Scientific Instrument Development and Application, University of Science and Technology of China, Hefei, 230026, China}

\author{Wei Wu}
\thanks{These authors contributed equally to this work.}
\affiliation{Laboratory of Spin Magnetic Resonance, School of Physical Sciences, Anhui Provincial Key Laboratory of Scientific Instrument Development and Application, University of Science and Technology of China, Hefei, 230026, China}
\author{Zihan Xie}
\affiliation{Laboratory of Spin Magnetic Resonance, School of Physical Sciences, Anhui Provincial Key Laboratory of Scientific Instrument Development and Application, University of Science and Technology of China, Hefei, 230026, China}
\affiliation{Hefei National Laboratory, University of Science and Technology of China, Hefei, 230088, China}
\author{Mengxiang Zhang}
\affiliation{Anhui Provincial Engineering Research Center for Quantum Precision Measurement, University of Science and Technology of China, Hefei 230088, China}
\author{Yi Li}
\affiliation{Laboratory of Spin Magnetic Resonance, School of Physical Sciences, Anhui Provincial Key Laboratory of Scientific Instrument Development and Application, University of Science and Technology of China, Hefei, 230026, China}
\affiliation{Hefei National Laboratory, University of Science and Technology of China, Hefei, 230088, China}

\affiliation{National Advanced Talent Cultivation Center for Physics,
University of Science and Technology of China, Hefei, 230026, China}
\author{Yue Li}
\affiliation{Laboratory of Spin Magnetic Resonance, School of Physical Sciences, Anhui Provincial Key Laboratory of Scientific Instrument Development and Application, University of Science and Technology of China, Hefei, 230026, China}
\author{Rixin Yao}
\affiliation{Laboratory of Spin Magnetic Resonance, School of Physical Sciences, Anhui Provincial Key Laboratory of Scientific Instrument Development and Application, University of Science and Technology of China, Hefei, 230026, China}
\affiliation{Hefei National Research Center for Physical Sciences at the Microscale, University of Science and Technology of China, Hefei, 230026, China}
\author{Yuqi Zhou}
\affiliation{Laboratory of Spin Magnetic Resonance, School of Physical Sciences, Anhui Provincial Key Laboratory of Scientific Instrument Development and Application, University of Science and Technology of China, Hefei, 230026, China}
\affiliation{Hefei National Laboratory, University of Science and Technology of China, Hefei, 230088, China}
\author{Xu Cheng}
\affiliation{Laboratory of Spin Magnetic Resonance, School of Physical Sciences, Anhui Provincial Key Laboratory of Scientific Instrument Development and Application, University of Science and Technology of China, Hefei, 230026, China}
\affiliation{Hefei National Laboratory, University of Science and Technology of China, Hefei, 230088, China}

\author{Han Pu}
\email{hpu@rice.edu}
\affiliation{Department of Physics and Astronomy,and Smalley-Curl Institute, Rice University, Houston, Texas 77251-1892, USA}

\author{Yiheng Lin}
\email{yiheng@ustc.edu.cn}
\affiliation{Laboratory of Spin Magnetic Resonance, School of Physical Sciences, Anhui Provincial Key Laboratory of Scientific Instrument Development and Application, University of Science and Technology of China, Hefei, 230026, China}
\affiliation{Hefei National Laboratory, University of Science and Technology of China, Hefei, 230088, China}
\affiliation{Hefei National Research Center for Physical Sciences at the Microscale, University of Science and Technology of China, Hefei, 230026, China}

	\begin{abstract}
        Temporal non-analyticities in the rate function of the Loschmidt echo manifests a class of dynamical quantum phase transitions (DQPTs) that has emerged as a powerful framework for understanding far-from-equilibrium many-body dynamics. 
        While such DQPT has been extensively studied theoretically in spin-boson systems such as the Dicke model, their experimental observation remains elusive. In particular, the dynamics of DQPT in asymmetric spin subspaces and under the influence of spin dissipation are largely unexplored. Here, we report an experimental study of temporal DQPT in a generalized Dicke model using a trapped-ion quantum simulator. By coupling a linear chain of $^{40}$Ca$^{+}$ ions to a collective center-of-mass motional mode, we probe the quench dynamics starting from both symmetric and asymmetric initial states. We extract the rate function and identify temporal turn-around points that are in quantitative agreement with theoretical predictions. Additionally, we investigate the impact of spin dissipation on these dynamics. Our results establish an experimental platform for probing complex many-body out-of-equilibrium phenomena and advance the development of hybrid oscillator–spin quantum simulators.        
        
	\end{abstract}
	
	\maketitle

\emph{Introduction.}
Understanding the non-equilibrium dynamics of quantum many-body systems is one of the central topics in modern physics~\cite{polkovnikov2011colloquium, eisert2015quantum, heyl2018dynamical}. 
The concept of dynamical quantum phase transitions (DQPT) extends the principles of equilibrium phase transitions to characterize real-time quantum many-body dynamics~\cite{heyl2013dynamical,heyl2018dynamical,jurcevic2017direct}. It encompasses two distinct classes of critical phenomena: The first type (DQPT-I) focuses on the behavior of the system's order parameter in the asymptotic steady state, where  critical values of Hamiltonian parameters separate distinct dynamical behaviours~\cite{sciolla2010quantum, klinder2015dynamical, zhiqiang2017nonequilibrium}. The second type (DQPT-II), which is the focus of this work, occurs in the transient time evolution following a quantum quench and is marked by non-analytic kinks in the Loschmidt echo rate function, which is the normalized logarithm of the overlap between the time-evolved state and the initial state~\cite{heyl2013dynamical, link2020dynamical}. 
While DQPT-II (hereafter DQPT) has been successfully observed in several pure spin systems, including the Ising~\cite{jurcevic2017direct}, Haldane~\cite{flaschner2018observation}, and Su-Schrieffer-Heeger~\cite{tian2019observation} models, its experimental realization in spin-boson systems remains elusive~\cite{link2020dynamical}.

The spin-boson systems such as the Dicke model~\cite{kirton2019introduction} which describes the coupling between collective spins and bosonic modes, provides a fundamental framework for understanding diverse critical phenomena. Examples include phonon-mediated electron attraction in superconducting materials and the down-conversion of light in photosynthesis~\cite{crane2024hybrid,giustino2017electron,ritz2002quantum}. Moreover, the open Dicke model has been proposed as a platform for observing DQPT~\cite{link2020dynamical}. Various experimental platforms have been utilized to quantum simulate Dicke models \cite{sun2025quantum,kaur2021spin,baumann2010dicke,chen2021experimental,wu2024experimental}. Among these, trapped ions provide an ideal platform for such simulations due to the readily engineered coupling between internal spins and quantized motional modes~\cite{leibfried2003quantum}. 
 While current implementations of Dicke-model dynamics~\cite{safavi2018verification,bullock2026quantum} are largely restricted to global control without single-site resolution, recent technical advances demonstrate the feasibility of site-resolved control and measurement~\cite{guo2026quantum,guo2024site}. Achieving such capabilities is crucial for exploring spin-boson physics beyond symmetric subspaces, including the dynamics of asymmetric spin sectors~\cite{mivehvar2024conventional,xu2024phase} and the effects of spin dissipation~\cite{boneberg2022quantum}. Therefore, experimental realization of the Dicke model with tunable spin-boson interactions, combined with individual addressing and readout, represents a key step toward probing rich out-of-equilibrium phenomena, including DQPT.

	In this work, we experimentally implement a generalized open Dicke model~\cite{kirton2019introduction} using a trapped-ion processor~\cite{leibfried2003quantum, haffner2008quantum}. The realization of spin-boson interactions is achieved by applying global laser fields to couple the internal pseudo-spin states of the ions to their center-of-mass (COM) collective motional modes~\cite{cirac1995quantum, molmer1999multiparticle, porras2004effective}, a technique that has been widely employed to engineer 
    quantum Rabi or Dicke dynamics~\cite{lamata2007dirac, pedernales2015quantum, safavi2018verification, cohn2018bang}. 
    We vary the system size from 4 to 8 ions and observe the Loshmidt echo rate function. 
    We also incorporate overlaps with all the other symmetric Dicke states alongside the initial state to further verify the constructed model.
    Going beyond symmetric Dicke subspace, we exploit flexible individual addressing and readout to prepare asymmetric initial spin states, where we also observe DQPT. 
    This could be helpful in studying subradiance, dark states \cite{gegg2018superradiant}, and the unconventional Dicke model with multistability and persistent oscillations \cite{mivehvar2024conventional}.
    We also study the influence of dissipation on DQPT~\cite{verstraete2009quantum, muller2012engineered}. By introducing controlled spin dissipation, we demonstrate that the signature of the DQPT remains robust against moderate levels of such operations~\cite{zunkovic2018dynamical}. 
	Our result contributes to benchmarking current and future hybrid oscillator-spin quantum simulators~\cite{crane2024hybrid}, opening up new avenues for investigating DQPT and general quantum dynamics in many-body systems~\cite{halimeh2025quantum,zache2019dynamical}. 

\emph{Implementation of the open Dicke model exhibiting DQPT}. We implement the open Dicke model \cite{kirton2019introduction} governed by the master equation $\dot{\rho} = -i[H, \rho] + \mathcal{L}(\rho)$. The coherent dynamics are dictated by the Hamiltonian
\begin{equation}
H = \Omega_c J_x + \frac{g}{2\sqrt{N}}(a J_- + a^\dagger J_+) + \delta_{s} J_z + \delta_{b} a^\dagger a, \label{eq:hamiltonian}
\end{equation}
where $a$ is the bosonic annihilation operator, and the collective spin operators for $N$ spins are $J_{k} = \frac{1}{2}\sum_{i=1}^N \sigma_{k}^{(i)}$ ($k = x, y, z$) and $J_\pm = \sum_{i=1}^N \sigma_{\pm}^{(i)}$. $\Omega_c$ and $g$ correspond to the strengths of the spin drive and spin-boson coupling. $\delta_{s}$ and $\delta_{b}$ denote spin and bosonic mode frequencies. The spin dissipation is captured by the dissipator $\mathcal{L}(\rho) = \sum_{k=1}^{N} ( L_k \rho {L_k}^\dagger - \frac{1}{2} \{ {L_k}^\dagger L_k, \rho \} )$ with jump operators $L_{k} = \sqrt{\gamma_s}\sigma^{(k)}_{-}$ for spin $k$, and $\gamma_s$ denotes the spin dissipation rate which is equal for all the spins.
    We define the rate function $r =-\frac{1}{N}\ln(L_s)= -\frac{1}{N} \ln\{\text{tr}[\rho_{s}(0)\rho_{s}(t)]\}$, with the return probability to the initial spin state $L_s = \text{tr}[\rho_{s}(0)\rho_{s}(t)]$ as the Loschmidt echo, where the spin density matrix $\rho_s = \mathrm{Tr}_{\mathrm{p}}(\rho)$ is obtained by taking the partial trace of the bosonic mode. The rate function is expected to exhibit cusps at certain times in the thermodynamic limit ($N \xrightarrow{}\infty$). These specific instants mark the occurrence of DQPT.
    According to Ref.~\cite{link2020dynamical}, the Loschmidt echo can be decomposed into two competing contributions $L_s = L_+ + L_-$, where $L_{\pm}=A e^{-NK_{\pm}}$, with $A>0$, $K_{\pm}>0$ acting as the potential function \cite{link2020dynamical}.  As $N \to \infty$, $r$ will approach $r_+ :=-\frac{1}{N}\ln(L_+)$ when $L_+ > L_-$, and $r_- :=-\frac{1}{N}\ln(L_-)$ when $L_+ < L_-$. Thus the non-analyticity emerging in the thermodynamic limit arises precisely at the crossings between $L_\pm$, i.e., where their magnitude switches. One typical decomposition \cite{link2020dynamical} adopted here is $L_{\pm} = \text{tr}[\rho_{s}(0)\rho_{s\pm}(t)]$, $\rho_{s\pm}(t) = \text{tr}_p[E_\pm \rho(t)]$, and
    $E_\pm = \int \frac{d^2\alpha}{\pi} |\alpha\rangle \langle\alpha| \Theta\left[\pm\text{Im}(e^{-i\theta_c}\alpha)\right]$ with $\ket{\alpha}$ the Bosonic coherent states, and $\theta_c$ determines the phase space dividing line.
    At finite $N$, the intersection points obtained for different $\theta_c$ are predominantly clustered around a specific point. As $N \to \infty$, this cluster of intersection points converges into a single point, which corresponds to the true critical point.
 There exists $\theta_c$ for which no crossing occurs between $r_+$ and $r_-$, which implies that one branch consistently represents the lower branch. As $N \to \infty$, this dominating lower branch inherently develops a non-analytic point at precisely the critical point, with  $r$ coinciding with it.
    In the following, the theoretically calculated $r_{\pm}$ and their intersections, combined with the experimental results, are used as evidence to support that the realized dynamics exhibit DQPT, as detailed in the Supplemental Material. Note that $E_{\pm}$ represents positive operator valued measurement (POVM) in bosonic modes and, in principle, will provide a means to measure two components separately and verify the occurrence of DQPT experimentally without going to the thermodynamic limit \cite{link2020dynamical}.  
	
	The experimental implementation of the above model employs a linear chain of $^{40}\text{Ca}^+$ ions, with $N$ ranging from 4 to 8. The qubit is encoded in the optical transition between the $|S_{1/2}, m = 1/2\rangle$ ($\left|\downarrow \right\rangle$, $\sigma_z\left|\downarrow\right\rangle = -\left|\downarrow\right\rangle$) ground state and the metastable $|D_{5/2}, m = 5/2\rangle$ ($\left|\uparrow \right\rangle$, $J_{+}\ket{\downarrow}=\ket{\uparrow}$) state. By applying a global laser field at 729~nm with on-resonance and sideband frequency tones, we couple these internal pseudo-spin states to the COM motional mode with a frequency of $\omega_z \approx2\pi\times 500\,\text{kHz}$. 
We initialize the system in a product state of spin and phonon. The initial phonon state is prepared by electromagnetic induced transparency (EIT) and sideband cooling sequence. The spin states can be prepared in  symmetirc $\left|\downarrow\downarrow\cdots\downarrow\right\rangle$ or asymmetric states, e.g., $\left|\downarrow\downarrow\cdots\uparrow\right\rangle$, by optical pumping and addressed single-spin rotation via a focused laser beam. After an evolution under $H$ for $t$, the spins are individually readout via site-resolved imaging using a camera. We then proceed to extract the return probability $L_s$. For instance, considering the initial state $\left|\downarrow\downarrow\downarrow\uparrow\right\rangle$, each single experimental shot yields a projective measurement result, identified by a pattern like ``bright-bright-bright-dark" and variations thereof, where dark (bright) corresponds to the $\left|\uparrow\right\rangle$ ($\left|\downarrow\right\rangle$) state of the corresponding ion. After repeating the experiment $5000$ times, we record the frequency $w$ of the ``bright-bright-bright-dark" outcome. The ratio $w/5000$ is then taken to be the return probability $L_s$,
from which we extract $r(t)$.  
    
    \emph{Dynamics in symemtric spin subspace.} We begin by considering the symmetric spin subspace in the absence of spin dissipation. Here $g=\eta_0 \times \Omega_b$, $\eta_0=0.155$ characterizes the spin-boson coupling (the Lamb-Dicke parameter in our setup). $\Omega_b=2\pi\times 10.5 ~\textrm{kHz}, \Omega_c=2\pi\times 1.6 ~\textrm{kHz}$ are realized by carrier and blue-sideband drivings. $\delta_s= -2\pi\times 2.2 ~\textrm{kHz}, \delta_b=2\pi\times 1.1 ~\textrm{kHz}$ are realized by detunings.
    The initial phonon states are measured to be thermal states with average phonon number $\bar{n} \approx \{0.96, 1.47, 1.45\}$ (see Supplemental Material for details). 
    We experimentally observe the evolution of $r$, as illustrated in Fig.~\ref{fig:2}, which agrees with the theoretical prediction. Furthermore, as the number of ions is increased from $4$ to $6$ and then to $8$, we observe a trend that is consistent with theoretical predictions. 
    The kink of the rate function is expected to approach a non-analytic point in the thermodynamic limit. As shown in the Supplemental Material, as $N$ increases (e.g., to $N=100$), $r$ closely tracks the lower branch  and the extremum of $r$ becomes sharper, clearly indicating the presence of DQPT in this model. The statistics of the intersection points with different $\theta_c$ are also plotted, with the mean indicated by the blue dashed line and the standard deviation represented by the blue shaded band. The intersections are obtained
by uniformly sampling $45$ values of $\theta_c$ from $-\pi/2$ to $\pi/2$. Most intersections are tightly clustered near the extremum, offering additional support for the presence of DQPT (see supplemental material for details). The spread of the intersections does not narrow with increasing ion number, this is mainly caused by the finite-size effect. In the supplemental material, we also measure overlaps with
all the other symmetric Dicke states for $N=4$. The agreement between theory and experiment provides a further verification of the implemented model. 
	
	\begin{figure}[t]
		\includegraphics[width=\columnwidth]{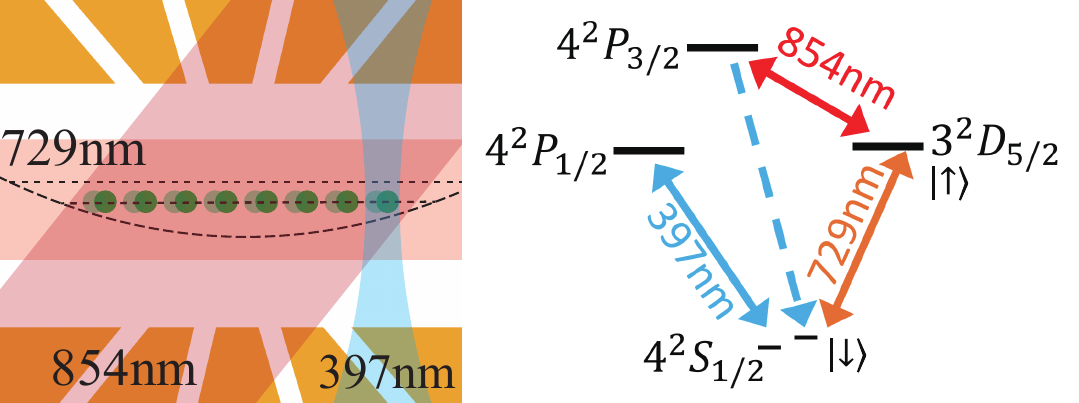}
		\caption{\label{fig:1}The trapped-ion setup and the level diagram for the implementation of the Dicke model. The Dicke model Hamiltonian is implemented using an axial 729~nm laser. Individual addressing and manipulation are achieved via radial 397 nm Raman beams. Spin dissipation is achieved by a global 854~nm beam.
        Site-resolved readout is performed using a  camera.}
	\end{figure}

\begin{figure}[t]
	\includegraphics[width=1\columnwidth]{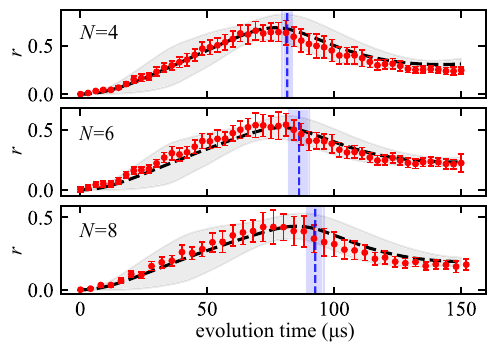}
	\caption{\label{fig:2} Theoretical (black dashed) and experimental (red) time evolution of the rate function for $N=4,6,8$. The initial phonon states are thermal states with $\bar{n} \approx \{0.96, 1.47, 1.45\}$. The gray band represents the error band accounting for experimental uncertainties. The mean of the intersections is indicated by the blue dashed line, and the blue band represents the $\pm 1\sigma$ interval ($\sigma = \{1.6,4.3,3.2\}$). }
\end{figure}

\begin{figure}[t]
		\includegraphics[width=\columnwidth]{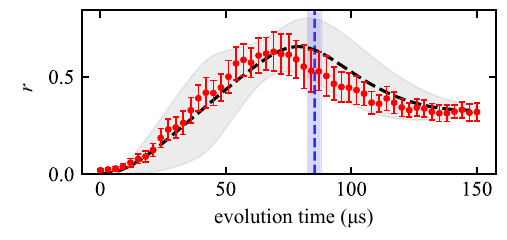}
		\caption{\label{fig:0001}  $r(t)$ starting from $|\psi_0\rangle = \left|\downarrow \downarrow \downarrow \uparrow\right\rangle$. The initial phonon state is a thermal state with $\bar{n} \approx 0.76$. The red dots represent the experimental data, and the black dashed line denotes theoretical predictions. The gray band indicates the error band, while the blue dashed line and blue band represent the mean of the intersections and the $\pm 1\sigma$ standard deviation interval ($\sigma = 2.8$), respectively. }
	\end{figure}

	\emph{Dynamics in asymmetric spin subspace.}
    Exploring asymmetric subspace allows us to obtain broader information regarding the constructed model and the associated DQPT. 
As an example, consider $N=4$ and initial spin state $\left|\downarrow \downarrow \downarrow \uparrow\right\rangle$, representing a state outside of the Dicke manifold. 
The experimental preparation of such state requires individual addressing capabilities. Starting from the spin state $\left|\downarrow \downarrow \downarrow \downarrow\right\rangle$  and an initial phonon thermal state with $\bar{n} \approx 0.76$ after optical pumping and cooling, we implement site-resolved spin flipping on the target ion by a 397~nm Raman addressing beam focused to the end of the ion chain. 
The results are shown in Fig.~\ref{fig:0001}, enabled by individual spin readout. Here $\eta_0=0.152$, with other parameters unchanged.
Despite of the lack of spin symmetry, by expressing the initial state as a superposition of states with fixed total angular momentum, efficient simulations (e.g., $N=100$) can still be achieved (see Supplemental Material). We find that $r$ adheres to the lower branch and the extremum of it sharpens, serving as a clear signature of the DQPT. This further demonstrates the generality of DQPT beyond symmetric subspace. The departure from the symmetric manifold paves the way for a comprehensive exploration of the full Hilbert space in future work. Specifically, it opens avenues to investigate the Dicke model with nonuniform couplings and its associated rich phenomena, including dark states, subradiance, and multicriticality \cite{mivehvar2024conventional,gegg2018superradiant,guo2026quantum, bullock2026quantum}.

\begin{figure}[t]
		\includegraphics[width=\columnwidth]{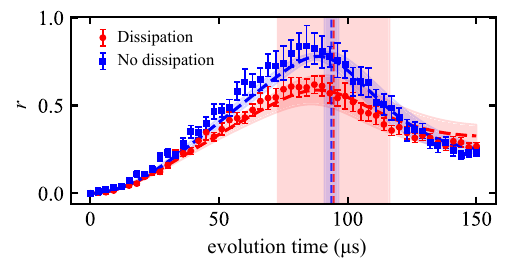}
		\caption{\label{fig:dis} Comparison of $r$ without (blue) and with (red) spin dissipation. The initial phonon state is a thermal state with $\bar{n} \approx 0.2$. For each case, the experimental data, theoretical predictions and error bands are represented by dots, dashed curves and shaded bands in the corresponding color, respectively. The mean of the intersections and the $\pm 1 \sigma$ standard deviation intervals ($\sigma=\{2.6,21.8\}$) are indicated by the vertical dashed lines and the shaded bands in the corresponding colors.} 
	\end{figure}

	\emph{Influence of applied dissipation on DQPT.}
	We now turn to investigate the behavior of DQPT under engineered spin dissipation.
   We add a global 854 nm dissipative beam to induce spin dissipation, as illustrated in Fig.~\ref{fig:1}. Here $\eta_0=0.128$, $N=4$, $\bar{n} \approx 0.2$, with other parameters unchanged. 
   The detuning of the 854 nm light from $|P_{3/2}, m=3/2\rangle$ is subject to experimental uncertainty, which introduces an additional, uncertain detuning in the effective dissipative two-level system (see Supplemental Material for details). 
   After accounting for this effect, the experimental data agrees with theoretical predictions, as demonstrated in Fig. \ref{fig:dis}. 
   The effective decay constant $\gamma_e$ is experimentally measured to be $2\pi \times (0.7 \pm 0.016) $ kHz and the effective detuning is fitted to be $\delta_e=2\pi \times (3.89 \pm 0.10)  \text{ kHz}$.
    By leveraging the permutation invariant property, efficient simulations (e.g., up to $N=25$) can be achieved (see Supplemental Material). We find that as $N$ increases, the extremum of $r$ migrates toward the $r_\pm$ intersection, $r$ also tracks the lower branch more closely, serving as a signature of the DQPT. This indicates the  robustness of DQPT behavior against moderate levels of dissipation and modification of the Hamiltonian parameters. Since spin dissipation also drives the system into the asymmetric subspace, this lays the foundation for future studies on the dissipation-induced leakage from the symmetric subspace and the associated dynamics \cite{boneberg2022quantum}.
	
	\emph{Conclusion.}
In summary, we report experimental studies of DQPT in an open Dicke model using a trapped-ion quantum simulator. We systematically investigate the DQPTs within the symmetric spin subspace, including finite-size scaling from 4 to 8 ions and the observation of overlaps with other symmetric Dicke states. The observed trend suggests the emergence of non-analytic behavior in the thermodynamic limit, which, combined with theoretical input, paves the way for verifying DQPT in finite-size systems. 
Leveraging our individual addressing and readout capabilities to access  asymmetric spin subspace, we observe persistent DQPT signatures, which further reflect the phenomenon's universality. Additionally,
we also investigated the impact of spin dissipation on DQPT. These pave the way for a comprehensive exploration of the full Hilbert space in future work. Crucially, accessing the non-symmetric sectors allows for the investigation of the Dicke model with nonuniform couplings and the rich phenomena such as subradiance and multi-criticality. 
These results benchmark the high controllability of our platform for studying coupled spin-boson system, 
lay the groundwork for future verifiable simulations of quantum many body dynamics, lattice gauge theories, and may facilitate the development of criticality-enhanced quantum sensing \cite{chu2021dynamic}.

\emph{Acknowledgment.}
\noindent We thank Xi-Wang Luo and Valentin Link for helpful discussion, and CIQTEK for technical support. This work was funded by the National Natural Science Foundation of China (Grant No.~92565306), the Quantum Science and Technology-National Science and Technology Major Project (Grant No.~2021ZD0301603), the Chinese Academy of Sciences (Grant No.~XDB1300000), and National Key Research and Development Program of China (Grant No.~2025YFE0217900). HP acknowledges support from the Welch Foundation (Grant No. C-1669).

	\bibliographystyle{apsrev4-2} 
    \bibliography{dpt}
	
\clearpage 
\onecolumngrid 

\begin{center}
    \textbf{\Large Supplemental Materials for: Temporal Dynamical Quantum Phase Transition in Dicke Model with Trapped Ions}\\[0.6cm]
    
    Ji Bian,$^{1,*}$ Wei Wu,$^{1,*}$ Zihan Xie,$^{1,2}$ Mengxiang Zhang,$^{3}$ Yi Li,$^{1,2,4}$ Yue Li,$^{1}$ Rixin Yao,$^{1,5}$ Yuqi Zhou,$^{1,2}$ Xu Cheng,$^{1,2}$ Han Pu,$^{6,\dagger}$ and Yiheng Lin$^{1,2,5,\ddagger}$\\[0.4cm]
    
    \small{
    $^1$\textit{Laboratory of Spin Magnetic Resonance, School of Physical Sciences, Anhui Provincial Key Laboratory of Scientific Instrument Development and Application, University of Science and Technology of China, Hefei, 230026, China}\\
    $^2$\textit{Hefei National Laboratory, University of Science and Technology of China, Hefei, 230088, China}\\
    $^3$\textit{Anhui Provincial Engineering Research Center for Quantum Precision Measurement, University of Science and Technology of China, Hefei 230088, China}\\
    $^4$\textit{National Advanced Talent Cultivation Center for Physics, University of Science and Technology of China, Hefei, 230026, China}\\
    $^5$\textit{Hefei National Research Center for Physical Sciences at the Microscale, University of Science and Technology of China, Hefei, 230026, China}\\
    $^6$\textit{Department of Physics and Astronomy, and Smalley-Curl Institute, Rice University, Houston, Texas 77251-1892, USA}\\
    $^*$\text{These authors contributed equally to this work.}\\
    $^\dagger$\text{hpu@rice.edu} \quad $^\ddagger$\text{yiheng@ustc.edu.cn}
    }
\end{center}
\vspace{0.8cm}

\twocolumngrid 

	\subsection{Verifying the Occurrence of DQPT}
We verify the existence of the DQPT as $N \to \infty$ as follows. As shown in Figs.~\labelcref{468n100,0001n100,dissn25} (corresponding to Figs.~\labelcref{fig:2,fig:0001,fig:dis} in the main text), with increasing $N$ (e.g., up to $N=100$), the extremum of $r$ sharpens and migrates toward the $r_\pm$ intersection, $r$ also tracks the lower branch more closely. These signify the formation of a non-analyticity in the thermodynamic limit. Note that while the choice of $\theta_c$ is arbitrary, certain values simply render this phenomenon more pronounced at finite $N$. There are values of $\theta_c$ where $r_+$ and $r_-$ do not cross, one branch consistently dictates the minimum. Nevertheless, as the system scales to $N \to \infty$, this dominant lower branch naturally exhibits a non-analytic kink exactly at the critical point. We thus conclude that the implemented model exhibits a DQPT.

\setcounter{figure}{0}
\renewcommand{\thefigure}{S\arabic{figure}}

\begin{figure}[h]
        \includegraphics[width=0.9\columnwidth]{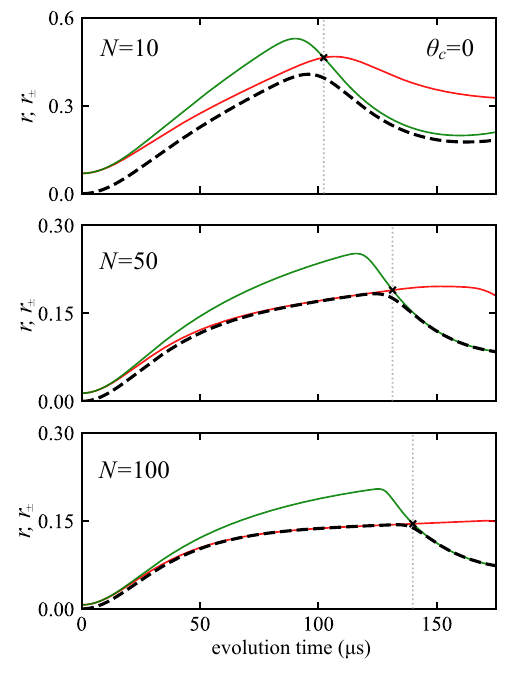}
        \caption{\label{468n100} Time evolution of $r_\pm$ and $r$ for increasing $N$, taking $\theta_c = 0$ as a representative example. The parameters are identical to those used in Fig.~\ref{fig:2} of the main text. The initial thermal phonon state is set to $\bar{n} = 1$ as an example. Similar results are also obtained for the no-dissipation case in Fig.~\ref{fig:dis}.}
    \end{figure}

\begin{figure}[h]
    \includegraphics[width=0.9\columnwidth]{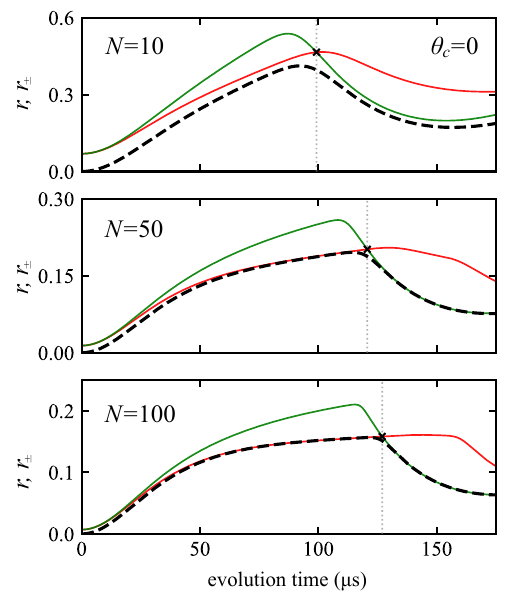}
    \caption{\label{0001n100} Time evolution of $r_\pm$ and $r$ for increasing $N$, taking $\theta_c = 0$ as a representative example.
    The parameters are identical to those in Fig.~\ref{fig:0001} of the main text.}
\end{figure}

\begin{figure}[h]
    \includegraphics[width=0.9\columnwidth]{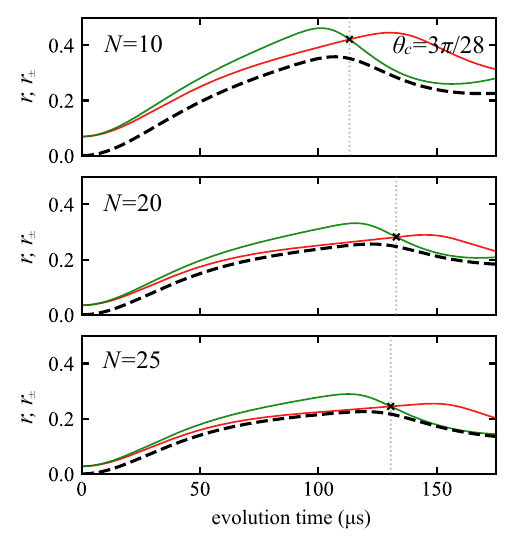}
    \caption{\label{dissn25} Time evolution of $r_\pm$ and $r$ for increasing $N$, taking $\theta_c = 3\pi/28$ as an example. The parameters are identical to the dissipative case shown in Fig.~\ref{fig:dis} of the main text.}
\end{figure}

\begin{figure}[htb]
        \includegraphics[width=0.9\columnwidth]{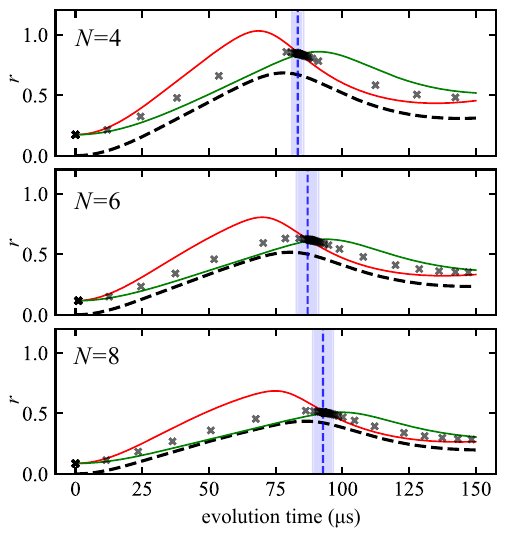}
        \caption{\label{fig:468scan} Theoretical $r$ (black dashed lines) corresponding to $N=4, 6, 8$ in the main text, and representative $r_+$ (red solid line) and $r_-$ (green solid line) at $\theta_c = -\pi/2$. Black crosses denote the intersections obtained by uniformly sampling 45 values of $\theta_c$ from $-\pi/2$ to $\pi/2$. The mean of the intersections is indicated by the blue dashed line, and the blue band represents the $\pm 1\sigma$ interval ($\sigma = \{1.6,4.3,3.2\}$). }
    \end{figure}

\begin{figure}[htb]
        \includegraphics[width=0.9\columnwidth]{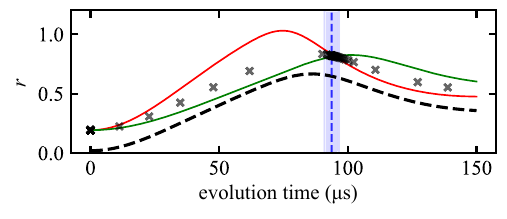}
        \caption{\label{fig:0001scan}  Theoretical $r$ (black dashed lines) corresponding to $N=4$, $|\psi_0\rangle = \left|\downarrow \downarrow \downarrow \uparrow\right\rangle$ in the main text, and representative $r_+$ (red solid line) and $r_-$ (green solid line) at $\theta_c = -\pi/2$. Black crosses denote the intersections obtained by uniformly sampling 45 values of $\theta_c$ from $-\pi/2$ to $\pi/2$. The mean of the intersections is indicated by the blue dashed line, and the blue band represents the $\pm 1\sigma$ interval ($\sigma = 2.8$). }
    \end{figure}

\begin{figure}[htb]
        \includegraphics[width=0.9\columnwidth]{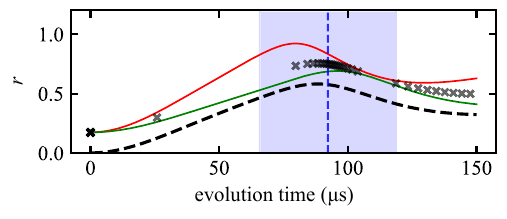}
        \caption{\label{fig:Dissipative_Scan}  Theoretical $r$ (black dashed lines) corresponding to $N=4$, with spin dissipation in the main text, and representative $r_+$ (red solid line) and $r_-$ (green solid line) at $\theta_c = -\pi/2$. Black crosses denote the intersections obtained by uniformly sampling 45 values of $\theta_c$ from $-\pi/2$ to $\pi/2$. The mean of the intersections is indicated by the blue dashed line, and the blue band represents the $\pm 1\sigma$ interval ($\sigma=21.8$). }
    \end{figure}

    The distribution of the intersections at finite $N$ provides additional support for the existence of a DQPT. At finite $N$, the intersections are not uniformly spread along the time axis; instead, they are predominantly distributed near the extremum of $r$.  In the thermodynamic limit, they will converge to a single DQPT point. Or, as explained above, no crossings will be present for some $\theta_c$. At finite sizes, this clustering effect of the crossings is already quite pronounced and can be correlated with the extremum points of $r$ from the experiment. As shown in Figs.~\labelcref{fig:468scan,fig:0001scan,fig:Dissipative_Scan}, the black crosses represent the intersections obtained by uniformly sampling 45 values of $\theta_c$ from $-\pi/2$ to $\pi/2$. The mean and $\pm1$ standard deviation interval of the crosses are represented by dashed blue lines and blue bands. They are predominantly distributed near the observed local extrema of $r$. 
    The intersection at $t=0$ is trivial, which arises from the symmetry of the initial phonon state in phase space, and is not taken into account in the statistical analysis. The imperfect coincidence between the extrema of $r$ and the mean of the $r_\pm$ intersections, as well as the finite spread of the intersection distribution, are caused by the finite-size effects. 
    Note the increase in the standard deviation with larger $N$ observed in Fig.~\ref{fig:468scan} is caused by the use of different initial phonon states for different $N$ (to match the experimental conditions), alongside the finite-size effects.
    This further demonstrates that our experimental observations are consistent with the systems exhibiting DQPT in the thermodynamic limit.


    \subsection{Overlaps with Different Dicke States}
\begin{figure}[htb]
		\includegraphics[width=\columnwidth]{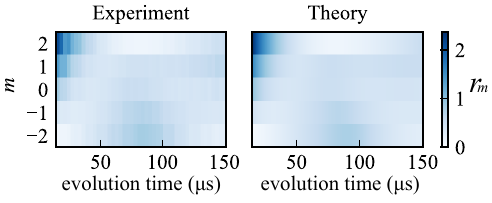} 
		\caption{\label{fig:overlaps} Experimental and theoretical generalized rate functions $r_m(t)$ for different collective spin states $|m\rangle$. }
	\end{figure}

     To further verify the implemented model, we observe the time evolution of additional observables.
    Specifically, we project the time-evolved state onto the basis of all the symmetric Dicke states with total angular momentum $J$, denoted by $|m\rangle$, satisfying $J_z|m\rangle = m|m\rangle$. Analogous to the Loschmidt echo, we define a generalized rate function:
	$
		r_m(t) = -\frac{1}{N} \textrm{ln}\{\textrm{tr}[ | m \rangle \langle m | \rho_s(t) ]\}.
	$	
	The experimental results are displayed in Fig.~\ref{fig:overlaps}. Here  $\eta_0=0.128$, $N=4$, $\bar{n}=0.20$ with other parameters unchanged. 
    $m \in \{-2, -1, 0, 1, 2\}$ for $J=N/2=2$. 
    The experimental results agree with theoretical predictions. 
    The manifestation of dynamical non-analyticities is not restricted to Loschmidt echo rate function~\cite{link2020dynamical}.
   They also presents in the rate functions defined by the overlaps with other symmetric Dicke states. Future theoretical and experimental works could explore DQPT under these generalized observables. 

	\subsection{Experimental Details}
	\emph{Implementation of the generalized Dicke model.}
	The experimental Hamiltonian is given by:
	\begin{equation}
		\begin{aligned}
			H &= \omega_0 J_z + \delta_p a^\dagger a \\
			&\quad + 2\Omega_b J_x \left(e^{i\omega_b t + i\eta(a+a^\dagger)} + \text{H.c.}\right) \\
			&\quad + 2\Omega_c J_x \left(e^{i\omega_0 t + i\eta(a+a^\dagger)} + \text{H.c.}\right),
		\end{aligned}
	\end{equation}
	where $\omega_b - \omega_0 = \delta_p + \delta_x$,  and $\eta=\eta_0/(2\sqrt{N})$.
	As $\eta \ll 1$, we expand the exponential term to the first order as $e^{i\eta(a+a^\dagger)} \approx 1 + i\eta(a+a^\dagger)$:
	\begin{equation}
		\begin{aligned}
			H &\approx \omega_0 J_z + \delta_p a^\dagger a \\
			&\quad + 2\Omega_b J_x \big[ (e^{i\omega_b t} + e^{-i\omega_b t}) \\
			&\quad\quad + i\eta(a+a^\dagger)(e^{i\omega_b t} - e^{-i\omega_b t}) \big] \\
			&\quad + 2\Omega_c J_x \big[ (e^{i\omega_0 t} + e^{-i\omega_0 t}) \\
			&\quad\quad + i\eta(a+a^\dagger)(e^{i\omega_0 t} - e^{-i\omega_0 t}) \big].
		\end{aligned}
	\end{equation}
	Define 
	$H_0 = \omega_0 J_z + (\delta_p + \delta_x)a^\dagger a$.
	Transforming into the rotating frame defined by $H_0$, 
	neglecting the fast-oscillating terms, we obtain:
	\begin{equation}
		\begin{aligned}
			H_{R,\text{RWA}} &= -\delta_x a^\dagger a + \Omega_c(J_+ + J_-) \\
			&\quad + \Omega_b \left(J_+ e^{-i(\delta_p+\delta_x)t} + J_- e^{i(\delta_p+\delta_x)t}\right) \\
			&\quad - i\eta \Omega_b (J_+ a^\dagger - J_- a).
		\end{aligned}
	\end{equation}
	When tuning the blue sideband transition we automatically introduce
	\begin{equation}
		\delta_x = -\frac{2\Omega_b^2}{\delta_p}.
	\end{equation}
	Thus we arrive at an time-independent effective Hamiltonian \cite{gamel2010time}:
	$
			H'_{\text{eff}} = -\delta_x a^\dagger a - \frac{2\Omega_b^2}{\delta_p + \delta_x} J_z 
			 + \Omega_c J_x - i\eta\Omega_b(J_+ a^\dagger - J_- a).
	$
With an extra transformation $H_{\text{eff}}=RH'_{\text{eff}}R^{\dagger}$, $R = \exp\left(i \frac{\pi}{2} a^\dagger a\right)$, we arrive at the final effective Hamiltonian 
$$H_{\text{eff}} = -\delta_x a^\dagger a - \frac{2\Omega_b^2}{\delta_p + \delta_x} J_z 
			+ \Omega_c J_x + \eta\Omega_b(J_+ a^\dagger + J_- a),$$
which is equivalent to Eq.\eqref{eq:hamiltonian}.
		
		\emph{Initial phonon state and heating rate}. The initial phonon state is prepared via a sequence of Doppler, electromagnetic-induced-transparency (EIT), and resolved-sideband cooling. Here we assume it to be a thermal equilibrium state.  
        By fitting the addressed blue-sideband oscillations \cite{cai2021observation,wineland1998experimental}, we reconstruct the initial phonon state as follows: 
        \begin{enumerate} 
        \item Following initial state preparation, measure the blue-sideband (BSB) oscillations after waiting times of 0, 100, 300, and 500 \unit{\micro\second}, respectively.
        \item For each measured curve, heating causes significant distortion at long evolution times, whereas the early-time evolution closely approximates a case where heating is negligible. Therefore, the first quarter of each curve is fitted using a thermal phonon distribution model, without a heating rate, to extract the approximate mean phonon number for different waiting times. A linear fit of these values then provides an initial estimate for the heating rate.
        \item Substitute this estimated heating rate into the master equation of the BSB oscillations to calculate the evolution of individual Fock states under heating. This generates a new set of basis curves, such that the BSB evolution curve of a mixed phonon state can be represented as a weighted sum of these basis.
        \item Extend the fitting range by an additional quarter and use the newly obtained basis curves to fit the data for different waiting times. This yields a new set of mean phonon numbers and a refined heating rate.
        \item Iteratively repeat steps 3 and 4 while progressively expanding the data window until the entire dataset is covered. Continue this process until the heating rate converges: specifically, until the difference of heating rates between consecutive estimates is less than half of the standard deviation of the heating-rate fit.

        \item Fit the BSB oscillation data with zero waiting time, using the above estimated heating rate and the corresponding basis curves.  Obtain the average phonon number of the initial phonon state.
        \end{enumerate}
        The heating rates are fitted to be $\{ 3009 \pm 349, 5547 \pm 947, 8415 \pm 1482, 1032 \pm 164\}$ phonons/s, corresponding to cases of $4$ ions in Fig.2 and $\left|\downarrow\downarrow\downarrow\uparrow \right\rangle$ (sharing the same heating rate data), $6$ ions, $8$ ions, spin dissipation and ifferent $m$ (sharing the same heating rate data), respectively. The corresponding initial average phonon numbers are fitted to be $\{0.96 \pm 0.12,
        0.76 \pm 0.05,
        1.47 \pm 0.14, 1.45 \pm 0.17, 0.20 \pm 0.03\}$, corresponding to cases of $4$ ions in Fig.2, asymmetric initial state $\left|\downarrow\downarrow\downarrow\uparrow \right\rangle$ (the BSB oscillation data at zero waiting time used to extract initial phonon state is separately measured), $6$ ions, $8$ ions, different $m$ and spin dissipation (sharing the same initial state), respectively.




    

		\emph{Initial state preparation and readout}. To prepare the $\left|\downarrow \downarrow \downarrow \uparrow\right\rangle$ initial state, we apply 397 nm addressing Raman pulses. In our current setup, the fidelity of these operations are limited by the intensity fluctuations of the Raman beams, resulting in an initial state preparation fidelity of approximately 98\% in this asymmetric-subspace experiment.



        \emph{Spin dissipation}. 
		Take $|P_{3/2}, m=3/2\rangle$ as an auxiliary state $|\textrm{aux}\rangle$,  $|\textrm{aux}\rangle$ decays back to $\left|\downarrow\right\rangle$ with a dissipation rate $\gamma_a= 2\pi \times 21.5$ MHz. Numerical simulations show that spontaneous emission to other states, along with its effect on the phonons, is negligible. Given that $\gamma_a $ is much larger than $ \text{other couplings}$ presented in the Hamiltonian, the excited state $|\textrm{aux}\rangle$ can be adiabatically eliminated. This reduces the system to an effective open two-level system governed by the master equation \cite{reiter2012effective}
		$$\begin{aligned} \dot{\rho} &= -i[H_e, \rho] + (L_e\rho L_e^\dagger -\frac{1}{2}\{L_e^\dagger L_e,\rho\}), \\ L_e &= \sqrt{\gamma_e}\left|\downarrow\right\rangle \left\langle \uparrow\right|, \end{aligned}$$
		where $H_e=H+\frac{\delta_e}{2}\sigma_z$,  which is accompanied by an induced effective detuning 
		$$\delta_e=-\frac{\Delta_0 \Omega_0^2}{4\Delta_0^2 + \gamma^2_a},$$ 
		where $\Delta_0$ and $\Omega_0$ are the 854 nm laser detuning and Rabi frequency,
		and the effective dissipation rate is 
		$$\gamma_e=\frac{\gamma_a \Omega_0^2}{4\Delta_0^2 + \gamma^2_a}.$$
		$\gamma_e$ is experimentally measured to be $2\pi \times (0.7 \pm 0.016) $ kHz. However, due to uncertainties in calibrating the 854 nm laser intensity and detuning, $\delta_e$ cannot be determined precisely. Consequently, we utilize the measured evolution of the population on $\left|\downarrow...\downarrow\right\rangle$ and $\left|\uparrow...\uparrow\right\rangle$ to calibrate $\delta_e$, finding that $\delta_e= 2\pi \times (3.89 \pm 0.10)   \text{ kHz}$. 

\end{document}